\documentstyle[12pt,epsfig]{article}

\begin{document}
\begin{titlepage}

\title{Gravitational Energy
of Kerr and Kerr Anti-de Sitter Space-Times in the
Teleparallel Geometry}
\author{J. F. da Rocha-Neto$\,^{a}$ and K. H.
Castello-Branco\,$^{b}$ \\
$^{a}$ Instituto de F\'{\i}sica Te\'{o}rica,
Universidade Estadual Paulista,\\
Rua Pamplona 145, 01405-900. S\~{a}o Paulo, Brazil.\\
$^{b}$ Instituto de F\'{\i}sica,
Universidade de S\~ao Paulo.\\
S\~ao Paulo, Brazil.
66318, 05315-970.\\}
\date{}
\maketitle

\begin{abstract}

In the context of the Hamiltonian formulation of the teleparallel
equivalent of general relativity we compute the gravitational
energy  of Kerr and Kerr Anti-de Sitter (Kerr-AdS) space-times.
The present calculation is carried out by means of an expression
for the energy of the gravitational field that naturally
arises from the integral form of the
constraint equations of the formalism. In each case, the energy
is exactly computed for finite and arbitrary spacelike
two-spheres,
without any restriction on the metric parameters. In particular,
we evaluate the energy at the outer event horizon of the
black holes.\\

\noindent $a$ E-mail: rocha@ift.unesp.br\,\,\,\,\,$b$ E-mail:
 karlucio@fma.if.usp.br\\
\end{abstract}
%\thispagestyle{empty}
%\vfill
\bigskip
%\end{titlepage}
\noindent {\bf I. Introduction}\par

\noindent

Teleparallel theories of gravity, whose basic entities are
tetrad fields $e_{a\mu}$ ( $a$ and $\mu$ are $SO(3,1)$ and space-time
indices, respectively) have been considered long time ago
by M\o ller \cite{Mol} in
connection with attempts to define the energy of the
gravitational field.
Teleparallel theories of gravity are defined on the
Weitzenb\"ock space-time \cite{Weit}, which is endowed with the
afinne connection
$\Gamma^\lambda_{\mu\nu}=e^{a\lambda} \partial_\mu
e_{a\nu}$
and where the curvature tensor, constructed out of this connection,
vanishes identically.
This connection defines a space-time with
an absolute parallelism or teleparallelism
of vector fields \cite{Schouten}. In this geometrical
framework the gravitational effects are due to the space-time
torsion corresponding to the above mentioned connection.

Although there exists an infinity of gravity theories in this
geometrical framework \cite{Hay}, here we will consider
only the teleparallel
equivalent of general relativity (TEGR) \cite{Hehl,Nes,Maluf1,
Per,Blago,Mielke}. The
TEGR is an alternative formulation of Einstein's
general relativity whose corresponding tetrad fields
satisfy the Einstein's equations in tetrad form.
As remarked by Hehl \cite{Hehl1}, by
considering Einstein's
general relativity as the best available alternative
theory of gravity, its teleparallel equivalent is the
next best one. Therefore it is
interesting to perform studies of the space-time
structure as described by the TEGR.

A simple definition for the energy of the gravitational field
has been established in the Hamiltonian formulation
of the TEGR \cite{Maluf1,Maluf2} in the framework of Schwinger's
time gauge condition \cite{Schwinger}. The
gravitational energy  is given by an
integral of a scalar density in the form of a total divergence
that appears in the Hamiltonian constraint of the theory.
The Hamiltonian formulation of the TEGR, with no {\it a priori}
restriction on the tetrad fields, has recently been
established \cite{Maluf3}.
Again, in this formulation,
an expression for the gravitational energy naturally arises \cite{MF},
in strict similarity with the procedure adopted in \cite{Maluf2}. In
this paper we apply this definiton of gravitational energy for the cases
of Kerr and Kerr-AdS space-times. The Kerr metric is one of the most
important known configurations of the gravitational field, consisting
in the only vacuum rotating black hole solution of Einstein's equations
\cite{Bicak}, whereas the Kerr-AdS metric is an important AdS
space. AdS spaces have received special attention in the current
literature due to the AdS/CFT correspondence \cite{Malda}. These
spaces have also been studied earlier, as for example, in
\cite{Hawk}, where it was showed that a (large) Schwarzschild-AdS black hole is
thermodynamically stable, whereas in \cite{Marc} the conserved charges
of AdS spaces were studied. Here we compute the gravitational energy of
Kerr and Kerr-AdS black holes enclosed by an arbitrary two-sphere of
radius  greater than or equal to their horizon radius $(r_+)$.
The knowledge of the distribution
of the gravitational energy for an arbitrary two-sphere of
radius $r\ge r_+$ is of importance in the description of other properties
of the black hole space-time \cite{AA}, such as an analysis of the
gravitational thermodynamic not restricted to the black hole event horizon.
In Ref. \cite{AA} a change in the area of black holes is related
to the fluxes of energy and angular momentum caried by gravitational
waves across the horizon of black holes.

Notation: space-time indices $\mu, \nu, ...$ and SO(3,1)
indices $a, b, ...$ run from 0 to 3. Time and space indices are
indicated according to
$\mu=0,i,\;\;a=(0),(i)$. The tetrad field $e^a\,_\mu$
yields the definition of the torsion tensor:
$T^a\,_{\mu \nu}=\partial_\mu e^a\,_\nu-\partial_\nu e^a\,_\mu$.
The flat, Minkowski space-time  metric is fixed by
$\eta_{ab}=e_{a\mu} e_{b\nu}g^{\mu\nu}= (-+++)$. \\

%%%%%%%%%%%%%%%%%%%%%%%%%%%%%%%%%%%%%%%%%%%%%%%%%%%%%%%%%%%%%%%
%%%%%%%%%%%%%%%%%%%%%%%%%%%%%%%%%%%%%%%%%%%%%%%%%%%%%%%%%%%%%%%

\noindent{\bf II. The Hamiltonian
Formulation of the TEGR}\par

The Hamiltonian
formulation of the TEGR developed in Ref. \cite{Maluf3},
without posing any a {\it priori} restriction on the
tetrad fields,
is obtained from the Lagrangian
density in empty space-time, given by
$$L(e)\;=\;-k\,e\,\left( {1\over 4} T^{abc}T_{abc} +
{1\over 2}T^{abc}T_{bac}-T^aT_a\right)
+ 2\partial_\mu\,(eT^{\mu}),\eqno(1)$$
\noindent where $e=det(e^a\,_\mu)$,
$T_{abc}=e_b\,^\mu e_c\,^\nu T_{a \mu \nu}$,
$T_b=T^a\,_{ab}$, $k={1\over 16\pi G}$ and $G$ is the
gravitational constant.
The Lagrangian $L$ is equivalent to that of Einstein-Hilbert
up to the four-divergence. However, in the cases where
boundary terms are taken into account, the surface term is
necessary to leave the action integral in an appropriate form
for general relativity \cite{Wald}.
By explicit calculations \cite{Maluf1,Blago} it is possible to show that
the variation of the action integral with respect to $e_{a\mu}$
yields the Einstein's equations in tetrad form
$${\delta L\over \delta e^{a\mu}} \equiv {k\over 2}e
\biggl\{R_{a\mu}(e) -  {1\over 2}e_{a\mu}R(e)\biggr\}.\eqno(2)$$
The Hamiltonian is obtained by the prescription $L=p\dot q -H$
and without making any kind of projection of metric
or tetrad variables to the three-dimensional spacelike hypersurface.
Dispensing with surface terms,
the total Hamiltonian density
reads \cite{Maluf3}
$$H(e_{ai},\Pi^{ai})
=e_{a0}C^a+\alpha_{ik}\Gamma^{ik}+\beta_k\Gamma^k\;.\eqno(3)$$
$\{C^a, \Gamma^k, \Gamma^{ik}\}$ is a set of first class constraints,
$\alpha_{ik}$ and $\beta_k$ are Lagrange  multipliers. More details are
given in Ref. \cite{Maluf3}.
The first term of the constraint $C^a$ is
given by a total divergence in the form
$C^a=-\partial_k \Pi^{ak}+Q^a$.
As in Ref. \cite{Maluf2} we identify this total
divergence on the three-dimensional spacelike hypersurface
as the energy-momentum density of the gravitational field
$$P^a=-\int_V d^3x\,\partial_i \Pi^{ai},\eqno(4)$$
\noindent where $V$ is an arbitrary spacelike volume. It is invariant
under coordinate transformations on the spacelike manifold, and
transforms as a vector under the global SO(3,1) group.
The definition above generalizes the analogous energy expression
(11) of the Ref. \cite{Maluf2} to tetrad fields
that are not restricted by the time gauge condition. If the time gauge
condition is imposed these expressions coincide \cite{Maluf3}.

The momenta $\Pi^{ak}$ are given by
$$\Pi^{ak}\;=\;k\,e\biggl\{
g^{00}(-g^{kj}T^a\,_{0j}-
e^{aj}T^k\,_{0j}+2e^{ak}T^j\,_{0j})$$
$$+g^{0k}(g^{0j}T^a\,_{0j}+e^{aj}T^0\,_{0j})
\,+e^{a0}(g^{0j}T^k\,_{0j}+g^{kj}T^0\,_{0j})$$
$$-2(e^{a0}g^{0k}T^j\,_{0j}+e^{ak}g^{0j}T^0\,_{0j})
-g^{0i}g^{kj}T^a\,_{ij}$$
$$+e^{ai}(g^{0j}T^k\,_{ij}-
g^{kj}T^0\,_{ij})-2(g^{0i}e^{ak}-g^{ik}e^{a0})
T^j\,_{ji} \biggr\}.\eqno(5)$$
\noindent For asymptotically flat space-times,
in the limit $r \rightarrow \infty$, taking
the $a=(0)$ component in Eq. (4) and integrating over
the whole three-dimensional spacelike hypersurface,
after a straightforward calculation, we arrive at \cite{MF}
$$P^{(0)}={1\over {16\pi G}}\int_{S\rightarrow \infty}dS_k(\partial_i
h_{ik}-\partial_k h_{ii})=E_{ADM},\eqno(6)\;$$
\noindent which is the ADM energy \cite{ADM}.
The above result motivates the definition of the gravitational
energy enclosed by  an arbitrary volume $V$ of the three-dimensional
spacelike as
$$E_{g} =-\int_{V} d^{3}x\;\partial_{i}(\Pi^{(0)i})
=-\int_S dS_i \Pi^{(0)i}.\eqno(7)$$
Such definition has led to consistent and relevant
results when applied to important configurations of the
gravitational field, such as the evaluation of the
irreducible mass of the Kerr black hole \cite{MF}.
Furthermore, it gives the Bondi mass in the appropriated limit when
the time gauge
is imposed \cite{MF}.
In fact, the definition (7) satisfies
all the conditions usually required for a quasilocal gravitational
energy expression.

%%%%%%%%%%%%%%%%%%%%%%%%%%%%%%%%%%%%%%%%%%%%%%%%%%%%%%%%%%%%%%%
\bigskip
\noindent {\bf III. Kerr and Kerr-AdS Space-times}\par

In Ref. \cite{MF} the energy of the Kerr black hole was
computed only to the two-sphere defined by $r=r_+$ (outer event horizon).
Differently of \cite{MF}, here we compute
the gravitational energy within the volume $V$ of an arbitrary
two-sphere of radius $r \ge r_+$ of a sapacelike hypersurface,
without any further restriction on the metric parameters.

In terms of Boyer-Lindquist coordinates the Kerr-AdS metric is given by
$$ds^2 = -{\Delta _r\over\rho^2}\biggl(dt - {a\over \chi}\sin^2
\theta d\phi\biggl)^2 + {\rho^2 \over \Delta _r}dr^2 + {\rho^2
\over \Delta _\theta}d \theta^2$$
$$+{\Delta_\theta\sin^2\theta\over\rho^2}\biggl[adt -
{(r^2+a^2)\over \chi}d\phi\biggr]^2,\eqno(8)$$
\noindent where
$$\Delta_r = (r^2+a^2)(1+r^2/l^2) - 2mr,\;\;\;\;
\Delta_\theta = 1 - a^2\cos^2\theta/l^2,$$
$$\chi = 1 - a^2/l^2,\;\;\;\;
\rho^2 = r^2 + a^2\cos^2\theta,$$
and $l$ is the AdS radius, related to the cosmological
constant by $\Lambda = -3/l^2$. The parameters $m$ and
$a$ are related to the mass and angular momentum of the
black hole, respectively.
In the limit $l\rightarrow \infty$ ($\Lambda = 0$), Eq.
(8) reduces to the Kerr asymptotically flat
solution.

The determination of tetrads that correspond to a given
metric has been carefully analysed recently in Sec. IV of \cite{MF}.
A set of tetrads associated to the metric (8) is

{\footnotesize
$$e_{a\mu}=
\pmatrix{-A&0&0&0\cr
B\sin\theta\sin\phi&C\sin\theta\cos\phi & Dr\cos\theta\cos\phi
& -Er\sin\theta\sin\phi\cr
-B\sin\theta\cos\phi& C\sin\theta \sin\phi & Dr\cos\theta\sin\phi
&  Er\sin\theta\cos\phi\cr
0&C\cos\theta & -Dr\sin\theta & 0\cr},\eqno(9)$$
}
\noindent where
$$A=\sqrt{{\bf A}+{{\bf B^2} \over {\bf C}}},\;\;\;
B=-{{\bf B}\over \sqrt{\bf C}\sin\theta},\;\;\;
C={\rho\over \sqrt{\Delta_r}},$$
$$D= {\rho\over r\sqrt{\Delta_\theta}}, \;\;\; E= {\sqrt{{\bf C}}
\over r\sin\theta},$$
\noindent and
$${\bf A}= {\Delta_r\over \rho^2} -
{a^2\Delta_\theta\sin^2\theta\over \rho^2},\;\;\;
{\bf B}= {a\sin^2\theta \over \rho^2\chi}\biggl(\Delta_r - (r^2+a^2)
\Delta_\theta\biggr),$$
$${\bf C}= {\sin^2\theta\over \rho^2\chi^2}\biggl(\Delta_\theta(r^2+a^2)^2 -
\Delta_r\sin^2\theta\biggr).$$
\noindent The components of the torsion tensor obtained out of the set
of tetrads above is presented in the appendix.

For an arbitrary spacelike volume $V$ difined by a
two-sphere the gravitational energy $E_g$ is
defined relationally, in the sense that it depends on the choice
of the reference space-time (see the discussion in Ref. \cite{MF}).
In the Kerr case, the set of tetrads above (for $\Lambda=0$) amounts to
choosing a {\it unique} reference space-time that is neither related
by a boost transformation nor rotating with respect to the physical space-time
(for a detailed discussion we refer the reader to \cite{MF}). For
the Kerr-AdS case, the above tetrad field (9) is probabily the most
suitable choice as well, since for $\Lambda=0$ it reduces to the
tetrad field associated to the Kerr metric, as can be easily verified.

In order to obtain the energy within a spherical surface ($r=const$) of
a three-dimensional spacelike hypersurface of the Kerr space-time, we
need to compute the component $\Pi^{(0)1}$ of $\Pi^{ak}$,
which is given in the appendix for the Kerr and
Kerr-AdS cases. After that,
we performe the integration (7) in the angular variables and, by means
of a long, but straightforward, calculation we achieve

$$E_g=-{1\over 4}\biggl[-\sqrt{r^2+a^2} + {r^2\over 2a}\ln \biggl(
{\sqrt{r^2+a^2}-a\over \sqrt{r^2+a^2}+a}\biggr)
+{i2\sqrt{\alpha}\over a}E\biggl(i{a\over r},\;\sqrt{
\Delta \over \alpha}\;r\biggr)$$

$$-{i\over a}\sqrt{
\Delta \over \alpha}\partial_r \alpha F\biggl(i{a\over r},\;
\sqrt{\Delta \over \alpha}\;r\biggr)
+{i\over a}\sqrt{\alpha \over \Delta}\partial_r \Delta
\biggl[F\biggl(i{a\over r},\;
\sqrt{\Delta\over \alpha}\;r\biggr) - E\biggl(i{a\over r},\;
\sqrt{\Delta \over \alpha}\;r\biggr)\biggr]\biggr],\eqno(10)$$

\noindent where $i^2 = -1$,
$$\alpha=(r^2+a^2)^2 - \Delta a^2,\;\;\; \Delta = r^2+a^2 - 2mr$$
%\noindent

\begin{figure}[ht]
\begin{center}
\hspace{-1.5cm} \epsfig{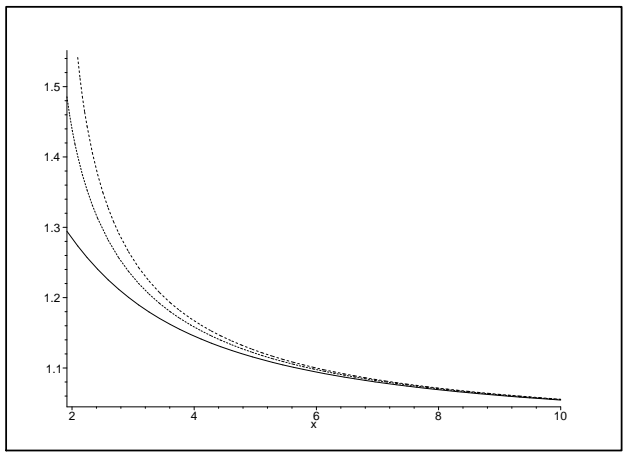} \vspace{0.0cm}
\caption{By using Eq. (10), here we plot $E_g/m$ versus $x=r/m$
for $a/m=0.4$ (doted), 0.7 (dashed), and 1.0 (solid) }
\end{center}
\end{figure}

\noindent and
$$E(x,z)=\int^{x}_{0}dy{\sqrt{1-z^2y^2}\over \sqrt{1-y^2}},$$
$$F(x,z)=\int^{x}_{0}dy{1\over \sqrt{1-y^2}\sqrt{1-z^2y^2}}$$
\noindent are the definitions of the elliptic functions $E(x,z)$

and $F(x,z)$, respectively. To arrive at the Eq. (10), the
integration in the angular variables was performed, in its final
part, using the Maple program. In Fig. 1 we plot the $E_g/m$
versus $r$ for some values of $a/m$. For large values of $r$ the
energy approaches of $m$, and in the horizon the energy approaches
of irreducible mass $M_{irr}$ of the black hole.

Let us now compare the results obtained from Eq. (10) with some
results in the literature.
In the asymptotic limit $r \rightarrow \infty$, the
gravitational energy given by (10) approaches the ADM energy $m$,
and for $a=0$, we obtain the energy of the Schwarzschild
black hole, in agreement with the result of
Brown and York \cite{Brown}.
In particular, for the surface defined
by $r=r_+$, differently of the Komar's integral \cite{Ko},
which gives the value $m$ \cite{Berg}, the above expression for the
gravitational energy $E_g$ reduces to
$$E_g(r_+) = m\left[{\sqrt{2p}\over 4} + {6p - k^2\over 4k}
\ln\left({\sqrt{2p} + k\over p}\right)\right],\eqno(11)$$
\noindent where $p=1+\sqrt{1-k^2},\;\;\;\; a=km,\;\;\;\;$ and $0\le k\le 1.$
The result (11) is the same obtained
in Ref. \cite{MF}, where $E_g(r_+)$ is compared with the irreducible
mass $2M_{irr}$ of the Kerr black hole, known from the work of Christodoulou
\cite{Chris}, as analysed in Ref. \cite{MF}. An excellent agreement
was found. This agreement is a crucial test for any local or
quasi-local gravitational energy expression.
In \cite{MF} it is shown a small difference
between (11) and $2M_{irr}$. However, it is argued in \cite{DM} that
the Martinez conjecture \cite{Mart} is not valid for large
values of angular momentum.

Recently, the gravitational energy of the Kerr black hole has been
analysed in the context of quasi-local energy by means
of the counterterms method \cite{DM}. In \cite{DM}, the
gravitational energy is computed in exact, closed analytic form only
at the surface defined by $r=r_+$. The result obtained in
\cite{DM} is different from that given here by Eq. (11)
(see Eq. (30) of the latter Ref.).
However, in the case of slow angular momentum, the Eq. (10) can be
expanded in powers of $a$ and the result obtained is
strictly the same obtained by Dehghani and Mann in \cite{DM}.
In Ref. \cite{Ag}, using pseudotensors, it is shown that the
energy of the Kerr black hole is $m$ which is confined to
its interior, because in that analysis the energy is independent
of the surface of integration. This result is the same
obtained in Ref. \cite{Berg} by using the Komar's integral.
%%%%%%%%%%%%%%%%%%%%%%%%%%%%%%%%%%%%%%%%%%%%%%%%%%%%%%%%%%%%

Let us now consider the gravitational energy for the
Kerr-AdS space-time. For space-times with different topologies
the appropriate gravitational action integrals require
a surface term that is specific to each topology.
Therefore, the correspoding Hamiltonian also acquires a surface term that
is determined by the topological boundary conditions \cite{Hawking1}.
Additional terms, such as the cosmological
constant, may appear in the Hamiltonian constraint.
Because of this, the action integral has an
extra term, as well as the Hamiltonian constraint.
Therefore, in the constraint $C^a$ there
appears an additional term given by $-2e^{a0}e\Lambda$.

As in the Kerr case, by means of a long but straightforward
calculation, we arrive at the gravitational energy
for the Kerr-AdS case contained within a surface of
constant radius $r$. Again, we perform the integration
of $\Pi^{(0)1}$ in terms of elliptic functions.
The integration is performed in exact, closed form for
arbitrary two-surfaces and values of the metric parameters.
The final expression for $E_g$ is

$$E_g= -{1\over 4}\biggl[-2\;{rl\over a}E\biggl({a\over l},\;i{l\over r}
\biggr) + i2\;{\sqrt{\alpha_l}\over\chi a}E\biggl(i{a\over r},\;
\sqrt{\beta_l\over \alpha_l}\;r\biggr)
-\;i{\sqrt{\Delta_r}\;l\;\partial_r\;\alpha_l\over a\chi\sqrt{\alpha_l(
r^2+l^2)}}F(i\eta,\;\zeta)$$

$$+\;i{\sqrt{\Delta_r}\;l^3\;\partial_r\;\beta_l\over\chi a\sqrt{
\alpha_l(r^2+a^2)}}\times
\biggl[-F(i\eta,\;\zeta)
+\; \Pi\biggl(i\eta,\;{r^2\over r^2 + l^2},\;
\zeta\biggr)\biggr]\biggr],\eqno(13)$$

\noindent where
$$\alpha_l= (r^2+a^2)^2 - \Delta_ra^2,\;\;\;
\beta_l= \Delta_r - {(r^2+a^2)^2\over l^2};$$
$$\eta=\sqrt{r^2+l^2\over l^2-a^2}\;{a\over r},\;\;\;\;\zeta=
\sqrt{\Delta_r(a^2+l^2)\over \alpha_l(r^2+l^2)}\;r$$
\noindent and
$$\Pi(x,y,k)=\int^{x}_{0}dz{1\over (1 - yz^2)\sqrt{1-z^2}
\sqrt{1 - k^2z^2}}\; $$
\noindent is the defination of the elliptic function $\Pi (x,y,k)$.

In the absence of angular momentum ($a=0$) it is easy
to check that (13) reduces to
$$E_g=r\biggl(1 - \sqrt{1-  {2m\over r} + {r^2\over l^2}}
\biggr),\eqno(14)$$
\noindent which is the energy within an arbitrary spacelike surface of
fixed radius $r$ in the Schwarschild-AdS space-time. The result (14)
has been obtained by Brown {\it et al} \cite{BCM}
in the context of the quasi-local energy, when the background subtraction
method is used, and
when the reference term is taken as $\epsilon_0(r)=-1/4\pi r$ ( in the
notation of Ref. \cite{BCM}). In the limit $r\rightarrow \infty$,
we find that (14) gives $E_g\rightarrow -\infty$. This result might
be expected,
since Anti-de Sitter space is a non-compact manifold with constant
negative curvature.

For the  especial case in which $E_g$ is evaluated
for the surface defined by $r=r_+$, i.e, for the external horizon
of the Kerr-AdS black hole, we find that (13) reduces to
$$E_g=-{1\over 4}\biggl[-{2lr_+\over a}E\biggl({a\over l},i
{l\over r_+}\biggr) + i{2(r^2_+ + a^2)\over\chi a}E\biggl(
i{a\over r_+},i{r_+\over l}\biggr)\biggr].\eqno(15)$$
\noindent This result differs from that obtained in Ref.
\cite{DM} in the context of the counterterms method. In the
latter reference the gravitational energy inside a volume
defined by the outer horizon is divergent in the limit $l
\rightarrow \infty$ (see Eq. (49) in \cite{DM}).
Differently of \cite{DM}, in this  limit,
expressions (13) and (15) reduces to those
of the Kerr case, given by (10) and (11), respectively.
This is a prominent result of the  energy expression (7),
and is consistent with the fact that the metric given
by Eq. (8) reduces to the Kerr metric in the
limit $l \rightarrow \infty$. This suggests that the flat
space limit of AdS space can, in fact, be defined as the limit
$l \rightarrow \infty$. Such a limit has been discussed, for example, in
\cite{Sussk} in the context of string theory. It is argued in \cite{Sussk}
that the non-perturbative matrix theory (M-Theory) in flat
space can be achieved by means of the AdS-CFT correspondence.
Then a holographic description of the flat space can be realized.
The gravitational energy for the Kerr-AdS case, in
the case of small angular momentum, can be expanded in powers
of $a$ and the result obtained is very similar to the
corresponding one obtained in \cite{DM}.

In Ref. \cite{Aros} conserved charges for spacetimes with local
Anti-de Sitter asymptotic geometry is proposed. The conserved
charges for Ker-AdS spacetime in Ref. \cite{Aros} are calculated
considering as a reference spacetime that is not the Minkowski
spacetime, but the Anti-de Sitter spacetime. So in this
case in the absence of angular momentum and mass the charges
are vanishes. On the other hand in our paper the background spacetime
is the Minkowski spacetime that is characterized by the total
absence of energy and for which our energy expression vanishes.
The result presented in \cite{Aros} in the limite $l \rightarrow
\infty$, differently from that presented here for the energy
expression, reduces to $m$ that is the same obtained with
Komar's charge.
Moreover, as remarked by Faddeev \cite{Faddeev}, an expression
for gravitational energy must vanish only in the absence of
matter and gravitational field; consequently when the metric
is that of the Minkowski spacetime.

%%%%%%%%%%%%%%%%%%%%%%%%%%%%%%%%%%%%%%%%%%%%%%%%%%%%%%%%%%%%%%%%
\bigskip

\noindent {\bf IV. Concluding remarks}\par

In this paper we have computed the gravitational
energy in the space-times of Kerr and Kerr-AdS black holes. In each
case the gravitational energy has been evaluated
exactly, in analytic closed form for
an arbitrary two-sphere of radius $r \ge r_+$, without any further restriction
on the metric parameters.
This is the major result of this work and constitutes
an advantage of the TEGR approach for the description
of the gravitational field energy as compared to
those based on the Hilbert-Einstein action integral,
where the energy has been calculated exactly only
in particular cases of the Kerr and Kerr-AdS
space-times (see, for example, \cite{DM}).
Our results for the gravitational energy
of the Kerr-AdS black hole reduce naturally to those of
Kerr in the limit $l \rightarrow \infty$
$(\Lambda = 0)$. This is a consistent result of the gravitational
energy expression (7) and
indicates that it is possible in fact to define the flat space
limit of the AdS space as $l \rightarrow \infty$.

Finally, in view of the above results we expect that the discussed
expression for the gravitational field
energy to be useful in the study of the
thermodynamics of self-graviting systems, particularly in AdS space-times.

\bigskip
\noindent{\bf Appendix: Calculation of torsion and momenta}\par
Now we present the nonvanish components of the torsion tensor
related to the set of tetrads given in Eq. (9). They are:
$$T_{(0)01} = \partial_r\; A,$$
$$T_{(0)02} = \partial_\theta\; A,$$
$$T_{(1)01} = \sin\phi\sin\theta\;\partial_r\;(B/\sqrt{C}),$$
$$T_{(1)02} = \sin\phi\;\partial_\theta\;(B/\sqrt{C}),$$
$$T_{(1)03} = \cos\phi (B/\sqrt{C}),$$
$$T_{(1)12} = \cos\theta\cos\phi\;\partial_r\;(\rho/\sqrt{\Delta_\theta})
 - cos\phi\;\partial_\theta\;(\sin\theta \rho/\sqrt{\Delta_r})$$
$$T_{(1)13} = \sin\theta\sin\phi(\rho/\sqrt{\Delta_r})
 - \sin\phi\;\partial_r\;\sqrt{C},$$
$$T_{(1)23} = \cos\theta\sin\phi(\rho/\sqrt{\Delta_\theta})
 - \sin\phi\;\partial_\theta\;\sqrt{C},$$
$$T_{(2)01} = -\cos\phi\;\partial_r\;(B/\sqrt{C}),$$
$$T_{(2)02} = -\cos\phi\;\partial_\theta\;(B/\sqrt{C}),$$
$$T_{(2)03} = \sin\theta (B/\sqrt{C}),$$
$$T_{(2)12} = \cos\theta\sin\phi\;\partial_r\;(\rho/\sqrt{\Delta_\theta})
 - \sin\phi\;\partial_\theta\;(\sin\theta \rho/\sqrt{\Delta_r}),$$
$$T_{(2)13} = \cos\phi\;\partial_r\;\sqrt{C}
 - \sin\theta\cos\phi(\rho/\sqrt{\Delta_r},$$
$$T_{(2)23} = \cos\theta\;\partial_\theta\;\sqrt{C}
 - \cos\theta\cos\phi(\rho/\sqrt{\Delta_\theta}),$$
$$T_{(3)12} = -\sin\theta\;\partial_r\;(\rho/\sqrt{\Delta_\theta})
 - \partial_\theta\;(\cos\theta \rho/\sqrt{\Delta_r}).$$
\noindent
From the above we can use Eq. (5) to show that $\Pi^{(0)1}$
for the Kerr-AdS case is given by
$$\Pi^{(0)1}=-2k\biggl[{\rho\sin\theta\over\sqrt{\Delta_\theta}} +
{\sin\theta\over\chi\rho}\biggl(\Delta_\theta(r^2+a^2)^2 -
\Delta_ra^2\sin^2\theta\biggr)^{1/2}$$

$$-{\sqrt{\Delta_r}\sin\theta\over \chi\rho\sqrt{\Delta_\theta}}
\partial_r\biggl(\Delta_\theta (r^2+a^2)^2 - \Delta_ra^2\sin^2
\theta\biggr)^{1/2}\biggr]\;.$$

\noindent For $l \rightarrow \infty \; (\Lambda =0)$, we obtain
$\Pi^{(0)1}$ corresponding to the Kerr case
$$\Pi^{(0)1}=-2k\,\sin\theta\biggl( \rho + {1\over \rho}
\sqrt{\alpha + \beta \cos ^2 \theta} - {\sqrt{\Delta}\over
\rho}\partial_{r}\sqrt{\alpha + \beta \cos ^2\theta} \biggr)\;.$$
\bigskip
\newpage
\noindent {\bf Acknowledgements}\par
\noindent This work was supported by FAPESP-Brazil. The authors
are grateful to J. W. Maluf for the critical reading of the manuscript
and comments. J. F. R-N would like to thank J. G. Pereira for the
hospitality at the Instituto de F\'{\i}sica
Te\'orica-IFT/UNESP-Brazil.

\end{titlepage}
\end{document}